\documentclass[aps,prl,twocolumn,groupedaddress,amsmath,superscriptaddress,floatfix,nofootinbib]{revtex4}

\usepackage{amsfonts}
\usepackage{amssymb}
\usepackage{amsmath}
\usepackage[dvips]{graphicx}
\usepackage{color}

\usepackage{amsmath}
\usepackage{graphicx}
\usepackage{amsfonts}
\usepackage{amssymb}
\usepackage{hyperref}
\usepackage{color}
\usepackage{mathrsfs}
\usepackage{isomath}
\usepackage{amsmath}
\usepackage{amsthm}
\usepackage{epstopdf}
\usepackage{txfonts}

\begin{document}

\title{Demonstration of Monogamy Relations for Einstein-Podolsky-Rosen Steering \\ in Gaussian
Cluster States}
\author{Xiaowei~Deng$^{\ddagger}$}
\address{State Key Laboratory of Quantum Optics and Quantum Optics Devices, Institute of Opto-Electronics, Shanxi University, Taiyuan 030006, China}
\address{Collaborative Innovation Center of Extreme Optics, Shanxi University, Taiyuan 030006, China}
\author{Yu~Xiang$^{\ddagger}$}
\address{Collaborative Innovation Center of Extreme Optics, Shanxi University, Taiyuan 030006, China}
\address{State Key Laboratory of Mesoscopic Physics, School of Physics, Peking University, Collaborative Innovation Center of Quantum Matter, Beijing 100871, China}
\author{Caixing~Tian}
\address{State Key Laboratory of Quantum Optics and Quantum Optics Devices, Institute of Opto-Electronics, Shanxi University, Taiyuan 030006, China}
\address{Collaborative Innovation Center of Extreme Optics, Shanxi University, Taiyuan 030006, China}
\author{Gerardo~Adesso}
\address{Centre for the Mathematics and Theoretical Physics of Quantum Non-Equilibrium Systems (CQNE), School of Mathematical Sciences, The University of Nottingham, Nottingham NG7 2RD, United Kingdom}
\author{Qiongyi~He}
\email{qiongyihe@pku.edu.cn}
\address{Collaborative Innovation Center of Extreme Optics, Shanxi University, Taiyuan 030006, China}
\address{State Key Laboratory of Mesoscopic Physics, School of Physics, Peking University, Collaborative Innovation Center of Quantum Matter, Beijing 100871, China}
\author{Qihuang~Gong}
\address{Collaborative Innovation Center of Extreme Optics, Shanxi University, Taiyuan 030006, China}
\address{State Key Laboratory of Mesoscopic Physics, School of Physics, Peking University, Collaborative Innovation Center of Quantum Matter, Beijing 100871, China}
\author{Xiaolong~Su}
\email{suxl@sxu.edu.cn}
\address{State Key Laboratory of Quantum Optics and Quantum Optics Devices, Institute of Opto-Electronics, Shanxi University, Taiyuan 030006, China}
\address{Collaborative Innovation Center of Extreme Optics, Shanxi University, Taiyuan 030006, China}
\author{Changde~Xie}
\address{State Key Laboratory of Quantum Optics and Quantum Optics Devices, Institute of Opto-Electronics, Shanxi University, Taiyuan 030006, China}
\address{Collaborative Innovation Center of Extreme Optics, Shanxi University, Taiyuan 030006, China}
\author{Kunchi~Peng}
\address{State Key Laboratory of Quantum Optics and Quantum Optics Devices, Institute of Opto-Electronics, Shanxi University, Taiyuan 030006, China}
\address{Collaborative Innovation Center of Extreme Optics, Shanxi University, Taiyuan 030006, China}

\begin{abstract}
Understanding how quantum resources can be quantified and distributed over
many parties has profound applications in quantum communication. As one of
the most intriguing features of quantum mechanics, Einstein-Podolsky-Rosen
(EPR) steering is a useful resource for secure quantum networks. By
reconstructing the covariance matrix of a continuous variable four-mode
square Gaussian cluster state subject to asymmetric loss, we quantify the
amount of bipartite steering with a variable number of modes per party, and
verify recently introduced monogamy relations for Gaussian steerability,
which establish quantitative constraints on the security of information
shared among different parties. We observe a very rich structure for the
steering distribution, and demonstrate one-way EPR steering of the cluster
state under Gaussian measurements, as well as one-to-multi-mode steering.
Our experiment paves the way for exploiting EPR steering in Gaussian cluster
states as a valuable resource for multiparty quantum information tasks.
\end{abstract}

\maketitle

Schr\"{o}dinger~\cite{Schrodinger35} put forward the term \textquotedblleft
steering\textquotedblright\ to describe the \textquotedblleft spooky
action-at-a-distance" phenomenon pointed out by Einstein, Podolsky, and
Rosen (EPR) in their famous paradox~\cite{EPR35, Reid89}. Wiseman, Jones,
and Doherty~\cite{Howard07PRL} rigorously defined the concept of steering in
terms of violations of local hidden state model, and revealed that steering
is an intermediate type of quantum correalation between entanglement~\cite%
{Schrodinger35ent,entRMP} and Bell nonlocality~\cite{Bell65,BrunnerRMP},
where local measurements on one subsystem can apparently adjust (steer) the
state of another distant subsystem~\cite%
{Howard07PRA,ReidRMP,Eric09,cavalcanti17review}. Such correlation is
intrinsically asymmetric with respect to the two subsystems~\cite%
{one-way-Theory, He15,
Adesso15,ReidJOSAB,OneWayNatPhot,OneWayPryde,OneWayGuo}, and allows
verification of shared entanglement even if the measurement devices of one
subsystem are untrusted~\cite{Eric09}. Due to this intriguing feature,
steering has been identified as a physical resource for one-sided
device-independent (1sDI) quantum cryptography~\cite%
{1sDIQKD,1sDIQKD_howard,HowardOptica,CV-QKDexp,prxresource}, secure quantum
teleportation~\cite{SQT13Reid,SQT15,SQT16_LiCM}, and subchannel
discrimination \cite{subchannel}.

Recently, experimental observation of multiparty EPR steering has been
reported in optical networks~\cite{ANUexp} and photonic qubits~\cite%
{Spainexp, USTCexp}. These experiments offer insights into understanding
whether and how this special type of quantum correlation can be distributed
over many different systems, a problem which has been recently studied
theoretically by deriving so-called \textit{monogamy relations}~\cite%
{ckw,Reidmonogamy, Kimmonogamy, GSmonogamy, Yumonogamy, Adesso16,
Shumingmonogamy}. It has been shown that the residual Gaussian steering
stemming from a monogamy inequality~\cite{Yumonogamy} can act as a
quantifier of genuine multipartite steering~\cite{genuine13} for pure
three-mode Gaussian states, and acquires an operational interpretation in
the context of a 1sDI quantum secret sharing protocol~\cite{GiannisQSS}.
However, beyond \cite{ANUexp}, no systematic experimental exploration of
monogamy constraints for EPR steering has been reported to date.

As generated via an Ising-type interaction, a cluster state features better
persistence of entanglement than that of a Greenberger-Horne-Zeilinger (GHZ)
state, hence is considered as a valuable resource for one-way quantum
computation~\cite%
{Raussendorf2001,Walther2005,Menicucci2006,vanLoockcluster,Gu2009} and
quantum communication~\cite{Teamwork,Mura,Zeng,Weedbrook}. Continuous
variable (CV) cluster states~\cite{Zhang2006,Loock2007}, which can be
generated deterministically, have been successfully produced for eight~\cite%
{Su2012}, 60~\cite{Chen} and up to 10,000 quantum modes~\cite{Yok2013}.
Several quantum logical operations based on prepared CV cluster states have
been experimentally demonstrated~\cite{Wang2010,Ukai2011,Ukai20112,Su2013}.
While the previous studies of multipartite steering mainly focus on the CV
GHZ-like states~\cite{VanLoockCVGHZ}, comparatively little is known about
EPR steering and its distribution according to monogamy constraints in CV
cluster states.

In this Letter, we experimentally investigate properties of bipartite
steering within a CV four-mode square Gaussian cluster state (see Fig.~\ref%
{fig:scheme}), and quantitatively test its monogamy relations~\cite%
{Reidmonogamy, Kimmonogamy, GSmonogamy, Yumonogamy, Adesso16}. By
reconstructing the covariance matrix of the cluster state, we measure the
quantifier of EPR steering under Gaussian measurements introduced in \cite%
{Adesso15}, for various bipartite splits. We find that the two- and
three-mode steering properties are determined by the geometric structure of
the cluster state. Interestingly, a given mode of the state can be steered
by its diagonal mode which is not directly coupled, but can not be steered
even by collaboration of its two nearest neighbors, although they are
coupled by direct interaction. These properties are different from those of
a CV four-mode GHZ-like state. We further present for the first time an
experimental observation of a `reverse' steerability, where the party being
steered comprises more than one mode. With this ability, we precisely
validate four types of monogamy relations recently proposed for Gaussian
steering (see Table~\ref{TableMono}) in the presence of loss~\cite%
{Reidmonogamy,Kimmonogamy, GSmonogamy,Yumonogamy,Adesso16}. Our study helps
quantify how steering can be distributed among different parties in cluster
states and link the amount of steering to the security of channels in a
communication network.

\begin{figure}[tbp]
\begin{center}
\includegraphics[width=85mm]{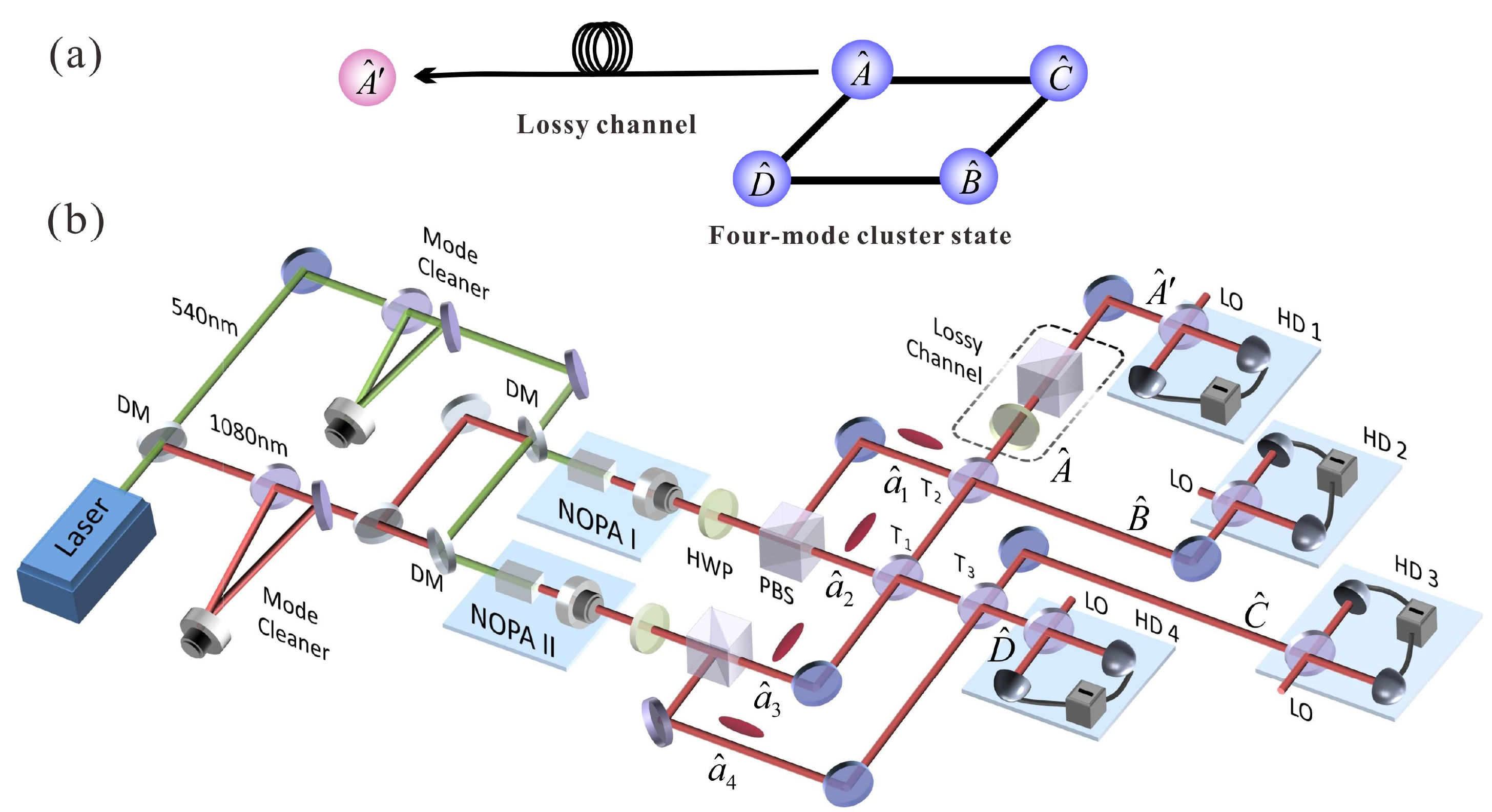}
\end{center}
\caption{Scheme of the experiment. (a) An optical mode ($\hat{A}$)\ of a
four-mode square cluster state is distributed over a lossy quantum channel.
(b) The experimental set-up. The squeezed states with $-3$ dB squeezing at
the sideband frequency of $3$ MHz are generated from two nondegenerate
optical parametric amplifiers (NOPAs). $T_{1}$, $T_{2}$ and $T_{3}$ are the
beam-splitters used to generate the cluster state. The lossy channel is
composed by a half-wave plate (HWP) and a polarization beam-splitter (PBS).
HD$_{1-4}$ denote homodyne detectors; LO denotes the local oscillator; and
DM denotes dichroic mirror.}
\label{fig:scheme}
\end{figure}

\begin{table}[b]
\begin{tabular}{llll}
\hline\hline
Type \  & Ref. \ \ \ \  & Inequality \qquad \qquad \qquad \qquad \  & 
Specifications \\ \hline
I & \cite{Reidmonogamy} & $\mathcal{G}^{A \rightarrow C} > 0 \ \ \Rightarrow
\ \ \mathcal{G}^{B \rightarrow C}=0$ & $n_A=n_B=n_C=1$ \\ 
II & \cite{Kimmonogamy,GSmonogamy} & $\mathcal{G}^{A \rightarrow C} > 0 \ \
\Rightarrow \ \ \mathcal{G}^{B \rightarrow C}=0$ & $n_A, n_B \geq 1$; $n_C=1$
\\ 
IIIa & \cite{Yumonogamy} & $\mathcal{G}^{C \rightarrow (AB)} - \mathcal{G}%
^{C \rightarrow A}- \mathcal{G}^{C \rightarrow B} \geq 0$ & $n_A=n_B=n_C=1$
\\ 
IIIb & \cite{Yumonogamy} & $\mathcal{G}^{(AB) \rightarrow C} - \mathcal{G}%
^{A \rightarrow C} - \mathcal{G}^{B \rightarrow C} \geq 0$ & $n_A=n_B=n_C=1$
\\ 
IVa & \cite{Adesso16} & $\mathcal{G}^{C \rightarrow (AB)} - \mathcal{G}^{C
\rightarrow A}- \mathcal{G}^{C \rightarrow B} \geq 0$ & $n_A, n_B, n_C \geq
1 $ \\ 
IVb & \cite{Adesso16} & $\mathcal{G}^{(AB) \rightarrow C} - \mathcal{G}^{A
\rightarrow C} - \mathcal{G}^{B \rightarrow C} \geq 0$ & $n_A, n_B \geq 1$; $%
n_C=1$ \\ \hline\hline
\end{tabular}%
\caption{Classification of monogamy relations for the bipartite quantifier $%
\mathcal{G}^{j\rightarrow k}$ of EPR steerability of party $k$ by party $j$
under Gaussian measurements, in a tripartite $(n_{A}+n_{B}+n_{C})$-mode
system $ABC$. Note: I $\sqsubseteq $ II and III $\sqsubseteq $ IV, where
\textquotedblleft $\sqsubseteq $\textquotedblright\ indicates being
generalized by; the relations in types II and IVb can be violated for $%
n_{C}>1$.}
\label{TableMono}
\end{table}

The CV cluster quadrature correlations (so-called nullifiers) can be
expressed by~\cite{Gu2009,Zhang2006,Loock2007} 
\begin{equation}  \label{cluster}
\big(\hat{p}_{a}-\sum_{b\in N_{a}}\hat{x}_{b}\big)\rightarrow 0,\qquad
\forall \quad a\in G
\end{equation}
where $\hat{x}_{a}=\hat{a}+\hat{a}^{\dagger }$ and $\hat{p}_{a}=(\hat{a}-%
\hat{a}^{\dagger })/i$ stand for amplitude and phase quadratures of an
optical mode $\hat{a}$, respectively. The modes of $a\in G$ denote the
vertices of the graph $G$, while the modes of $b\in N_{a}$ are the nearest
neighbors of mode $\hat{a}$. For an ideal cluster state the left-hand side
of Eq.~(\ref{cluster}) tends to zero, so that the state is a simultaneous
zero eigenstate of these quadrature combinations in the limit of infinite
squeezing~\cite{Gu2009}.

As a unit of two-dimensional cluster state, a four-mode square cluster state
as shown in Fig.~\ref{fig:scheme}(a) can be used to establish a quantum
network~\cite{GiannisQSS,shenPRA}. The cluster state of the optical field is
prepared by coupling two phase-squeezed and two amplitude-squeezed states of
light on an optical beam-splitter network, which consists of three optical
beam-splitters with transmittance of $T_{1}=1/5$ and $T_{2}=T_{3}=1/2$,
respectively, as shown in Fig.~\ref{fig:scheme}(b)~\cite{Supp}. We
distribute mode $\hat{A}$ of the state in a lossy channel [Fig.~\ref%
{fig:scheme}(a)]. The output mode is given by $\hat{A}^{\prime }=\sqrt{\eta }%
\hat{A}+\sqrt{1-\eta }\hat{\upsilon}$, where $\eta $ and $\hat{\upsilon}$\
represent the transmission efficiency of the quantum channel and the vacuum
mode induced by loss into the quantum channel, respectively.

The properties of a ($n_A+m_B$)-mode Gaussian state $\rho _{AB}$ of a
bipartite system can be determined by its covariance matrix 
\begin{equation}  \label{eq:CM}
\sigma _{AB}=\left( 
\begin{array}{cc}
A & C \\ 
C^{\top } & B%
\end{array}
\right) ,
\end{equation}
with elements $\sigma _{ij}=\langle \hat{\xi}_{i}\hat{\xi}_{j}+\hat{\xi}_{j} 
\hat{\xi}_{i}\rangle /2-\langle \hat{\xi}_{i}\rangle \langle \hat{\xi}%
_{j}\rangle $, where $\hat{\xi}\equiv (\hat{x}_{1}^{A},\hat{p}_{1}^{A},...,%
\hat{x}_{n}^{A},\hat{p}_{n}^{A},\hat{x}_{1}^{B},\hat{p}_{1}^{B},...,\hat{x}%
_{m}^{B},\hat{p}_{m}^{B})$ is the vector of the amplitude and phase
quadratures of optical modes. The submatrices $A$ and $B$ are corresponding
to the reduced states of Alice's and Bob's subsystems, respectively. The
partially reconstructed covariance matrix $\sigma _{A^{\prime }BCD}$, which
corresponds to the distributed mode $\hat{A}^{\prime }$ and modes $\hat{B}$, 
$\hat{C}$ and $\hat{D}$, is measured by four homodyne detectors~\cite%
{Supp,Steinlechner}.

The steerability of Bob by Alice ($A\rightarrow B$) for a ($n_A+m_B$)-mode
Gaussian state can be quantified by~\cite{Adesso15} 
\begin{equation}  \label{eqn:parameter}
\mathcal{G}^{A\rightarrow B}(\sigma _{AB})=\max \left\{0, \underset{j:\bar{%
\nu}_{j}^{AB\backslash A}<1}{-\sum }\ln (\bar{\nu}_{j}^{AB\backslash
A})\right\},
\end{equation}
where $\bar{\nu}_{j}^{AB\backslash A}$ $(j=1,...,m_B)$ are the symplectic
eigenvalues of $\bar{\sigma}_{AB\backslash A}=B-C^{\mathsf{T}}A^{-1}C$,
derived from the Schur complement of $A$ in\ the covariance matrix $\sigma
_{AB}$. The quantity $\mathcal{G}^{A\rightarrow B}$ is a monotone under
Gaussian local operations and classical communication \cite{Adesso16} and
vanishes iff the state described by $\sigma _{AB}$ is nonsteerable by
Gaussian measurements \cite{Adesso15}. The steerability of Alice by Bob [$%
\mathcal{G}^{B\rightarrow A}(\sigma _{AB})$] can be obtained by swapping the
roles of $A$ and $B$.

\begin{figure}[tbp]
\begin{center}
\includegraphics[width=85mm]{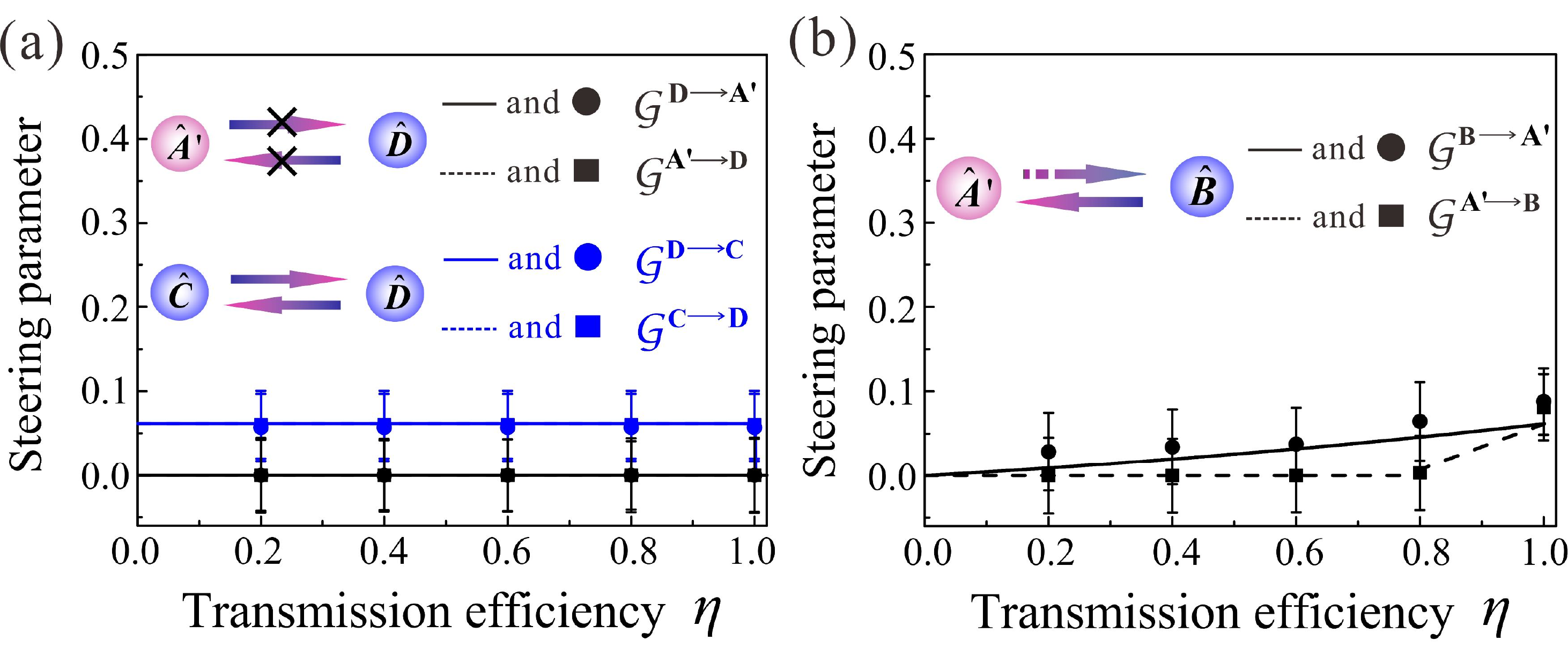}
\end{center}
\caption{Gaussian EPR steering between two modes of the cluster state. (a)
There is no EPR steering between neighboring modes $\hat{A}^{\prime }$ and $%
\hat{D}$ under Gaussian measurements, while diagonal modes $\hat{C}$ and $%
\hat{D}$ can steer each other with equal power. (b) One-way EPR steering
between modes $\hat{A}^{\prime }$ and $\hat{B}$ under Gaussian measurements.
Additional ($1+1$)-mode partitions are shown in Fig.~S2 in \protect\cite%
{Supp}. In all the panels, the quantities plotted are dimensionless. The
lines and curves represent theoretical predictions based on the theoretical
covariance matrix as calculated in \protect\cite{Supp}. The dots and squares
represent the experimental data measured at different transmission
efficiencies. Error bars represent $\pm $ one standard deviation and are
obtained based on the statistics of the measured noise variances.}
\label{fig:2modes}
\end{figure}

Figure~\ref{fig:2modes} shows a selection of results for the steerability
between any two modes [i.e., ($1+1$)-mode partitions] of the cluster state
under Gaussian measurements. Surprisingly, as shown in Fig.~\ref{fig:2modes}%
(a) and Fig.~S2 in \cite{Supp}, we find that steering does not exist between
any two neighboring modes, as one might have expected due to the direct
coupling as shown in the definition of cluster state in Eq.~(\ref{cluster}).
Instead, two-mode steering is present between diagonal modes which are not
directly coupled, as shown in Fig.~\ref{fig:2modes}. This observation can be
understood as a consequence of the monogamy relation (type-I) derived from
the two-observable ($\hat{x}$ and $\hat{p}$) EPR criterion ~\cite%
{Reidmonogamy}: two distinct modes cannot steer a third mode simultaneously
by Gaussian measurements. In fact, as shown in Fig.~\ref{fig:scheme}, mode $%
\hat{C}$ and mode $\hat{D}$ are completely symmetric in the cluster state.
Thus, if $\hat{A}^{\prime }$ could be steered by $\hat{C}$, it should be
equally steered by $\hat{D}$ too, which, on the contrary, is forbidden by
the type-I monogamy relation. However, there is no such constraint for mode $%
\hat{B}$. As a comparison, in a CV GHZ-like state, pairwise steering is
strictly forbidden between any two modes based upon the same argument as the
state is fully symmetric under mode permutations~\cite{ckw, MengGHZ}. Thus,
we conclude that a cluster state features richer steerability properties,
due to the inherent asymmetry induced by its geometric configuration.

We further investigate quantitatively the robustness of the two-mode
steering when transmission loss is imposed on one of the two parties. In
Fig.~\ref{fig:2modes}(b), we show the steering parameter defined in Eq.~(\ref%
{eqn:parameter}) by varying the transmission efficiency $\eta $ of the lossy
channel. When the lossy mode $\hat{A}^{\prime }$ is the steered party, we
find that the non-lossy steering party $\hat{B}$ can always steer $\hat{%
A^{\prime }}$, although the steerability is reduced with increasing loss.
However, the presence of loss plays a vital role if $\hat{A}^{\prime }$ is
the steering party. In fact, if the transmission efficiency $\eta $ is lower
than a critical value of $\sim 0.772$, the Gaussian steering of $\hat{A}%
^{\prime }$ upon $\hat{B}$ is completely destroyed. This leads to a
manifestation of \textquotedblleft one-way" steering within the region of $%
\eta \in (0,0.772)$, as previously noted in other types of entangled states~%
\cite{OneWayNatPhot,OneWayPryde,OneWayGuo, ANUexp}. However, we remark that
in our experiment we are limited to Gaussian measurements for the steering
party, which leaves open the possibility that $A^{\prime }\rightarrow B$
steering could still be demonstrated for smaller values of $\eta $ by
resorting to suitable non-Gaussian measurements \cite{OneWayPryde,NhaSciRep}.

\begin{figure}[tbp]
\begin{center}
\includegraphics[width=85mm]{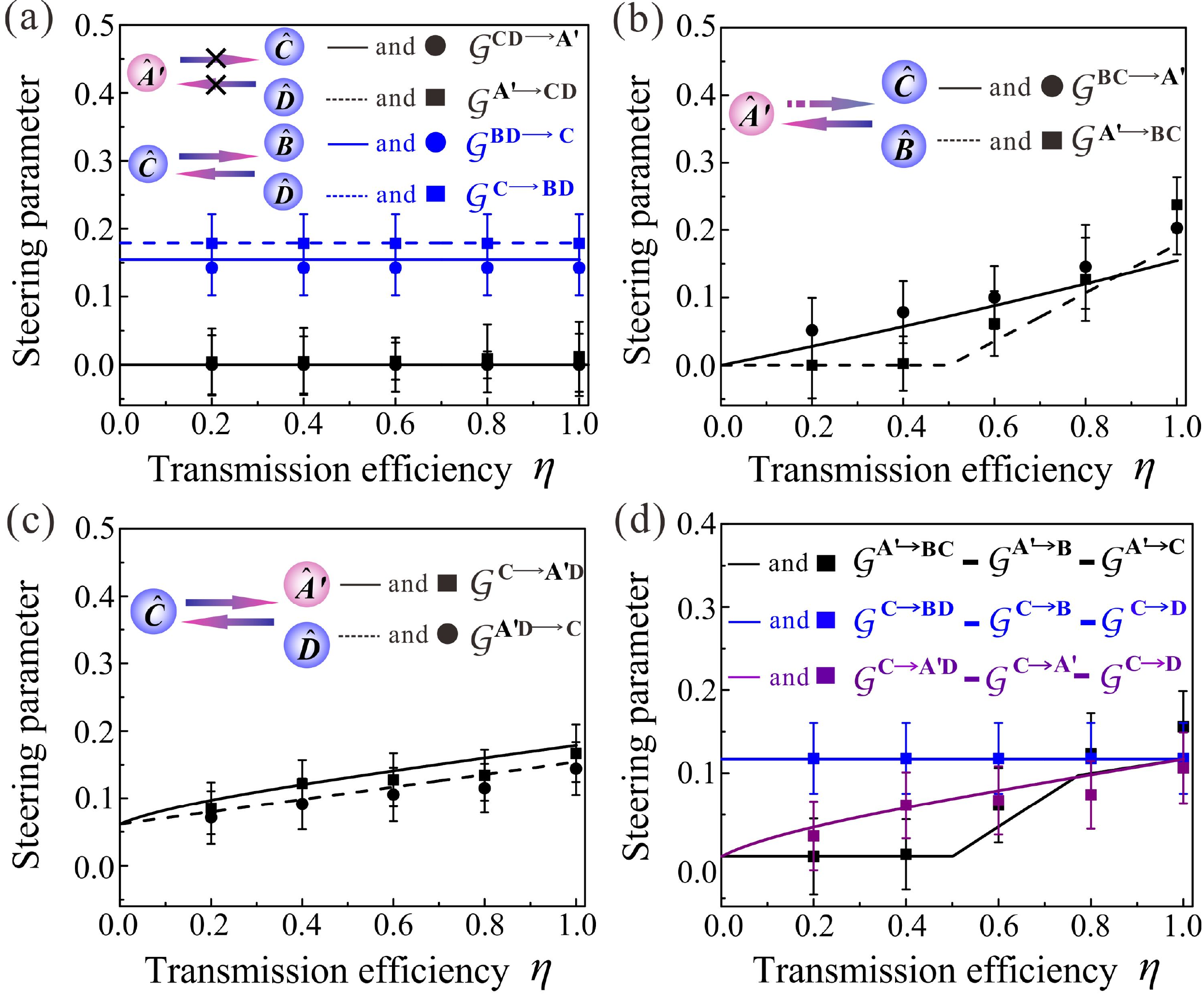}
\end{center}
\caption{Gaussian EPR steering between one and two modes of the cluster
state. (a) Mode $\hat{A}^{\prime }$ cannot be steered by the collaboration
of two nearest neighboring modes \{$\hat{C}$,$\hat{D}$\} even though they
are directly coupled; while $\hat{C}$ and \{$\hat{B}$,$\hat{D}\}$ can steer
each other. (b) One-way EPR steering between modes $\hat{A}^{\prime }$ and \{%
$\hat{B}$, $\hat{C}$\} under Gaussian measurements. (c) $\hat{C}$ and \{$%
\hat{A}^{\prime }$,$\hat{D}$\} can steer each other asymmetrically and the
steerability grows with increasing transmission efficiency, reflecting the
different effect when loss happens on steering or steered channel. (d)
Validation of CKW-type monogamy for steering (type-III). Additional
partitions are shown in Fig.~S3 in \protect\cite{Supp}. In all the panels,
the quantities plotted are dimensionless. The lines and curves represent
theoretical predictions based on the theoretical covariance matrix as
calculated in \protect\cite{Supp}. The dots and squares represent the
experimental data measured at different transmission efficiencies. Error
bars represent $\pm $ one standard deviation and are obtained based on the
statistics of the measured noise variances.}
\label{fig:3modes}
\end{figure}

Since mode $\hat{A}^{\prime }$ is coupled to its two nearest neighbors $\hat{%
C}$ and $\hat{D}$ on each side, one may wonder whether the two neighboring
modes can jointly steer $\hat{A}^{\prime }$. Figures~\ref{fig:3modes} and S3
in \cite{Supp} show the steerability between one mode and any two other
modes of the cluster state [i.e., ($1+2$)-mode and ($2+1$)-mode partitions]
under Gaussian measurements. Interestingly, we find that mode $\hat{A}%
^{\prime }$ still cannot be steered even by the collaboration of modes $\hat{%
C}$ and $\hat{D}$ ($\mathcal{G}^{CD\rightarrow A^{\prime }}=0$) [Fig.~\ref%
{fig:3modes}(a)], but can be steered so long as the diagonal mode $\hat{B}$
is involved ($\mathcal{G}^{BC\rightarrow A^{\prime }}=\mathcal{G}%
^{BD\rightarrow A^{\prime }}>0$) [Fig.~\ref{fig:3modes}(b)]. This phenomenon
is determined unambiguously from a generalized monogamy relation applicable
to the case of the steering party consisting of an arbitrary number of modes
(type-II)~\cite{Kimmonogamy, GSmonogamy}. As mode $\hat{B}$ can always steer 
$\hat{A}^{\prime }$ [shown in Fig.~\ref{fig:2modes}(b)], the other group \{$%
\hat{C}$, $\hat{D}$\} is forbidden to steer the same mode simultaneously. We
stress that this property is again in stark contrast to the case of CV
four-mode GHZ-like state, where any two modes \{$\hat{\imath},\hat{\jmath}$%
\} can collectively steer another mode $\hat{k}$ \cite{MengGHZ} as there is
no two-mode steering to rule out this possibility. Similarly, mode $\hat{C}$
can only be steered by a group comprising the diagonal mode $\hat{D}$ [$%
\mathcal{G}^{BD\rightarrow C}>0$ shown in Fig.~\ref{fig:3modes}(a), and $%
\mathcal{G}^{A^{\prime }D\rightarrow C}>0$ shown in Fig.~\ref{fig:3modes}%
(c)]. We also show that the collective steerability $\mathcal{G}%
^{BC(D)\rightarrow A^{\prime }}$ [solid curve in Fig.~\ref{fig:3modes}(b)]
is significantly higher than the steerability by $\hat{B}$ mode alone $%
\mathcal{G}^{B\rightarrow A^{\prime }}$ [solid curve in Fig.~\ref{fig:2modes}%
(b)], suggesting that although the neighboring modes $\hat{C}$ and $\hat{D}$
cannot steer $\hat{A}$ by themselves, their roles in assisting collective
steering with mode $\hat{B}$ are non-trivial.

We further measure, for the first time, the steerability when the steered
party comprises more than one mode, i.e., steering parameters of $(1+2)$%
-mode configurations, which are shown in Fig.~\ref{fig:3modes} and in
Fig.~S3 in \cite{Supp}. The loss imposed on $\hat{A}$ also leads to
asymmetric steerability $\mathcal{G}^{BC\rightarrow A^{\prime }}\neq 
\mathcal{G}^{A^{\prime }\rightarrow BC}$, and a parameter window for one-way
steering (under the restriction of Gaussian measurements) with $\eta \in
(0,0.5]$, as shown in Fig.~\ref{fig:3modes}(b). In addition, our results $%
\mathcal{G}^{D\rightarrow BC}>0$ [$\mathcal{G}^{D\rightarrow BC}=\mathcal{G}%
^{C\rightarrow BD}$, Fig.~\ref{fig:3modes}(a)] and $\mathcal{G}^{A^{\prime
}\rightarrow BC}>0$ when $\eta >0.5$ [Fig.~\ref{fig:3modes}(b)] also confirm
experimentally that, when the steered system is composed of at least two
modes, it can be steered by more than one party simultaneously, i.e., the
type-II monogamy relation is lifted \cite{GSmonogamy}.

Using the results of $(1+2)$-mode steerability, we also present the first
experimental examination of the type-III monogamy relation, called
Coffman-Kundu-Wootters (CKW)-type monogamy in reference to the seminal study
on monogamy of entanglement \cite{ckw}, which quantifies how the steering is
distributed among different subsystems~\cite{Yumonogamy}. For a three-mode
scenario, the CKW-type monogamy relation reads 
\begin{equation}
\mathcal{G}^{k\rightarrow (i,j)}(\sigma _{ijk})-\mathcal{G}^{k\rightarrow
i}(\sigma _{ijk})-\mathcal{G}^{k\rightarrow j}(\sigma _{ijk})\geq 0,
\end{equation}%
where $i,j,k\in \{\hat{A}^{\prime },\hat{B},\hat{C},\hat{D}\}$ in our case.
We have experimentally verified that this monogamy relation is valid for all
possible types of $(1+2)$-mode steering configurations; some of them are
shown in Fig.~\ref{fig:3modes}(d).

\begin{figure}[tbh]
\begin{center}
\includegraphics[width=85mm]{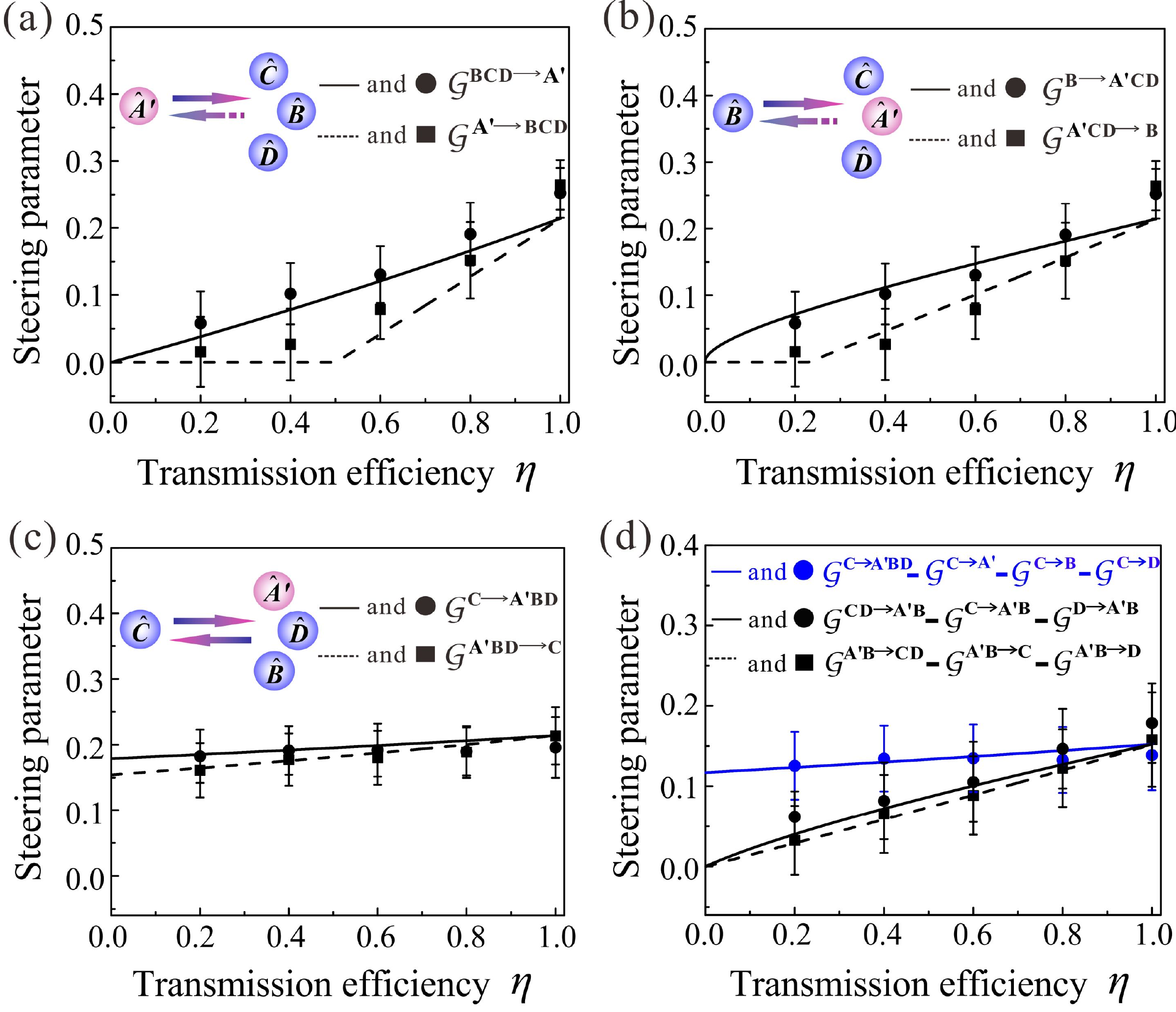}
\end{center}
\caption{Gaussian EPR steering between one and three modes in the cluster
state. (a) One-way EPR steering under Gaussian measurements between modes $%
\hat{A}^{\prime }$ and $\{\hat{B}$, $\hat{C}$, $\hat{D}\}$ with directional
property. (b) One-way EPR steering under Gaussian measurements between modes 
$\hat{B}$ and $\{\hat{A}^{\prime }$, $\hat{C}$, $\hat{D}\}$. (c) Asymmetric
steering between modes $\hat{C}$ and $\{\hat{A}^{\prime }$, $\hat{B}$, $\hat{%
D}\}$. (d) Monogamy of steering quantifier for $(1+3)$- and $(2+2)$-mode
partitions. In all the panels, the quantities plotted are dimensionless. The
lines and curves represent theoretical predictions based on the theoretical
covariance matrix as calculated in \protect\cite{Supp}. The dots and squares
represent the experimental data measured at different transmission
efficiencies. Error bars represent $\pm $ one standard deviation and are
obtained based on the statistics of the measured noise variances.}
\label{fig:4modes}
\end{figure}

Next, we study the steerability between one and the remaining three modes
within the cluster state, i.e., ($1+3$)- and $(3+1)$-mode partitions. As
shown in Figs.~\ref{fig:4modes}(a), (b), one-way EPR steering (under
Gaussian measurements) is observed for bipartitions $(\hat{A}^{\prime }+\hat{%
B}\hat{C}\hat{D})$ and $(\hat{B}+\hat{A}^{\prime }\hat{C}\hat{D})$ when $%
\eta \leq 0.5$ and $\eta \leq 0.228$, respectively. The asymmetry between
the two steering directions for the bipartition $(\hat{C}+\hat{A}^{\prime }%
\hat{B}\hat{D})$ grows with increasing transmission efficiency, but no
one-way property is observed in this case [Fig.~\ref{fig:4modes}(c)], since
mode $\hat{C}$ and mode $\hat{D}$ can always steer each other independently.
Quantitatively, the ($1+3$)- and $(3+1)$-mode steerability degrees are
further enhanced in comparison to the ($1+2$) and $(2+1)$ mode cases, even
when the newly added mode alone cannot steer or be steered by the other
party. We also confirm that the generalized CKW-type monogamy inequality $%
\mathcal{G}^{k\rightarrow (i,j,l)}-\mathcal{G}^{k\rightarrow i}-\mathcal{G}%
^{k\rightarrow j}-\mathcal{G}^{k\rightarrow l}\geq 0$ holds in this
four-mode scenario, as shown in Fig.~\ref{fig:4modes}(d).

Finally, our experiment also validates for the first time general monogamy
inequalities for Gaussian steerability with an arbitrary number of modes per
party (type-IV)~\cite{Adesso16}. As a typical example of $(2+2)$-mode
steering, our experimental results demonstrate that the steerability of $(%
\hat{A}^{\prime }\hat{B}+\hat{C}\hat{D})$-mode partitions satisfies the
following inequalities 
\begin{subequations}
\label{eq:monogamy}
\begin{eqnarray}
&&\mathcal{G}^{A^{\prime }B\rightarrow CD}-\mathcal{G}^{A^{\prime
}B\rightarrow C}-\mathcal{G}^{A^{\prime }B\rightarrow D}\geq 0,
\label{eq:monogamya} \\
&&\mathcal{G}^{CD\rightarrow A^{\prime }B}-\mathcal{G}^{C\rightarrow
A^{\prime }B}-\mathcal{G}^{D\rightarrow A^{\prime }B}\geq 0,
\label{eq:monogamyb}
\end{eqnarray}%
as indicated in Fig.~\ref{fig:4modes}(d). We have verified that both these
monogamy relations are also valid for all possible $(2+2)$-mode
configurations in this cluster state. Note that, in general, Eq.~(\ref%
{eq:monogamyb}) can be violated on other classes of states \cite{Adesso16}.

In summary, the structure and sharing of EPR steering distributed over two-,
three-, and four-mode partitions have been demonstrated and investigated
quantitatively for a CV four-mode square Gaussian cluster state subject to
asymmetric loss. By generating the cluster state deterministically and
reconstructing its covariance matrix, we obtain a full steering
characterization for all bipartite configurations. For general cases with
arbitrary numbers of modes in each party, we quantify the bipartite
steerability by Gaussian measurements, and provide experimental confirmation
for four types of monogamy relations which bound the distribution of
steerability among different modes, as summarized in Table~\ref{TableMono}.
Even though our state does not display genuine multipartite steering \cite%
{genuine13}, several innovative features are observed, including the
steerability of a group of two or three modes by a single mode, and the fact
that a given mode of the state can be steered by its diagonal mode which is
not directly coupled, but can not be jointly steered by its two directly
coupled nearest neighbors.

Our work thus provides a concrete in-depth understanding of EPR steering and
its monogamy in paradigmatic multipartite states such as cluster states. 
In turn, this can be useful to gauge the usefulness of these states for
quantum communication technologies. For instance, secure CV teleportation
with fidelity exceeding the no-cloning threshold requires two-way Gaussian
steering \cite{SQT15}, which arises in various partitions in our state,
e.g.~between $\hat{A}^{\prime }$ and $\hat B$ for sufficiently large
transmission efficiency [see Fig.~\ref{fig:2modes}(b)]. Furthermore, the
amount of Gaussian steering directly bounds the secure key rate in CV 1sDI
quantum key distribution and secret sharing \cite%
{HowardOptica,Yumonogamy,GiannisQSS}. Combined with a stronger initial
squeezing level, the techniques used here could be adapted to demonstrate
these protocols among many sites over lossy quantum channels.

\medskip

This research was supported by National Natural Science Foundation of China
(Grants No. 11522433, No. 11622428, No. 61475092, and No. 61475006),
Ministry of Science and Technology of China (Grants No. 2016YFA0301402 and
No. 2016YFA0301302), X. Su thanks the program of Youth Sanjin Scholar, Q. He
thanks the Cheung Kong Scholars Programme (Youth) of China, GA thanks the
European Research Council (ERC) Starting Grant GQCOP (Grant No.~637352) and
the Foundational Questions Institute (fqxi.org) Physics of the Observer
Programme (Grant No.~FQXi-RFP-1601). \newline
$^{\ddagger}$X. Deng and Y. Xiang contributed equally to this work.

\bigskip 

\appendix

\begin{center}
{\large \bfseries Supplemental Material}
\end{center}

\setcounter{equation}{0} \setcounter{figure}{0} \renewcommand\thefigure{S%
\arabic{figure}} \renewcommand\theequation{S\arabic{equation}}

\section{Details of the experimental setup}

In the experiment, the $\hat{x}$-squeezed and $\hat{p}$-squeezed states are
produced by non-degenerate optical parametric amplifiers (NOPAs) pumped by a
common laser source, which is a continuous wave intracavity
frequency-doubled and frequency-stabilized Nd:YAP-LBO (Nd-doped YAlO$_{3}$
perorskite-lithium triborate) laser. Two mode cleaners are inserted between
the laser source and the NOPAs to filter noise and higher order spatial
modes of the laser beams at 540 nm and 1080 nm, respectively. The
fundamental wave at 1080 nm wavelength is used for the injected signals of
NOPAs and the local oscillators of homodyne detectors. The second-harmonic
wave at 540 nm wavelength serves as the pump field of the NOPAs, in which
through an intracavity frequency-down-conversion process a pair of signal
and idler modes with the identical frequency at 1080 nm and the orthogonal
polarizations are generated.

Each of NOPAs consists of an $\alpha $-cut type-II KTiOPO4 (KTP) crystal and
a concave mirror. The front face of KTP crystal is coated to be used for the
input coupler and the concave mirror serves as the output coupler of
squeezed states, which is mounted on a piezo-electric transducer for locking
actively the cavity length of NOPAs on resonance with the injected signal at 
$1080$ nm. The transmissivities of the front face of KTP crystal at 540 nm
and 1080 nm are $21.2\%$ and $0.04\%$, respectively. The end-face of KTP is
cut to $1^{\circ }$ along y-z plane of the crystal and is antireflection
coated for both 1080 nm and 540 nm \cite{Zhou}. The transmissivities of
output coupler at 540 nm and 1080 nm are $0.5\%$ and $12.5\%$, respectively.
In our experiment, all NOPAs are operated at the parametric deamplification
situation \cite{Zhou,Su2007}. Under this condition, the coupled modes at $%
+45^{\circ }$ and $-45^{\circ }$ polarization directions are the $\hat{x}$%
-squeezed and $\hat{p}$-squeezed states, respectively \cite{Su2007}. The
quantum efficiency of the photodiodes used in the homodyne detectors are
95\%. The interference efficiency on all beam-splitters are about 99\%.

\section{Preparation and verification of the square cluster state}

\begin{figure}[tbp]
\begin{center}
\includegraphics[width=85mm]{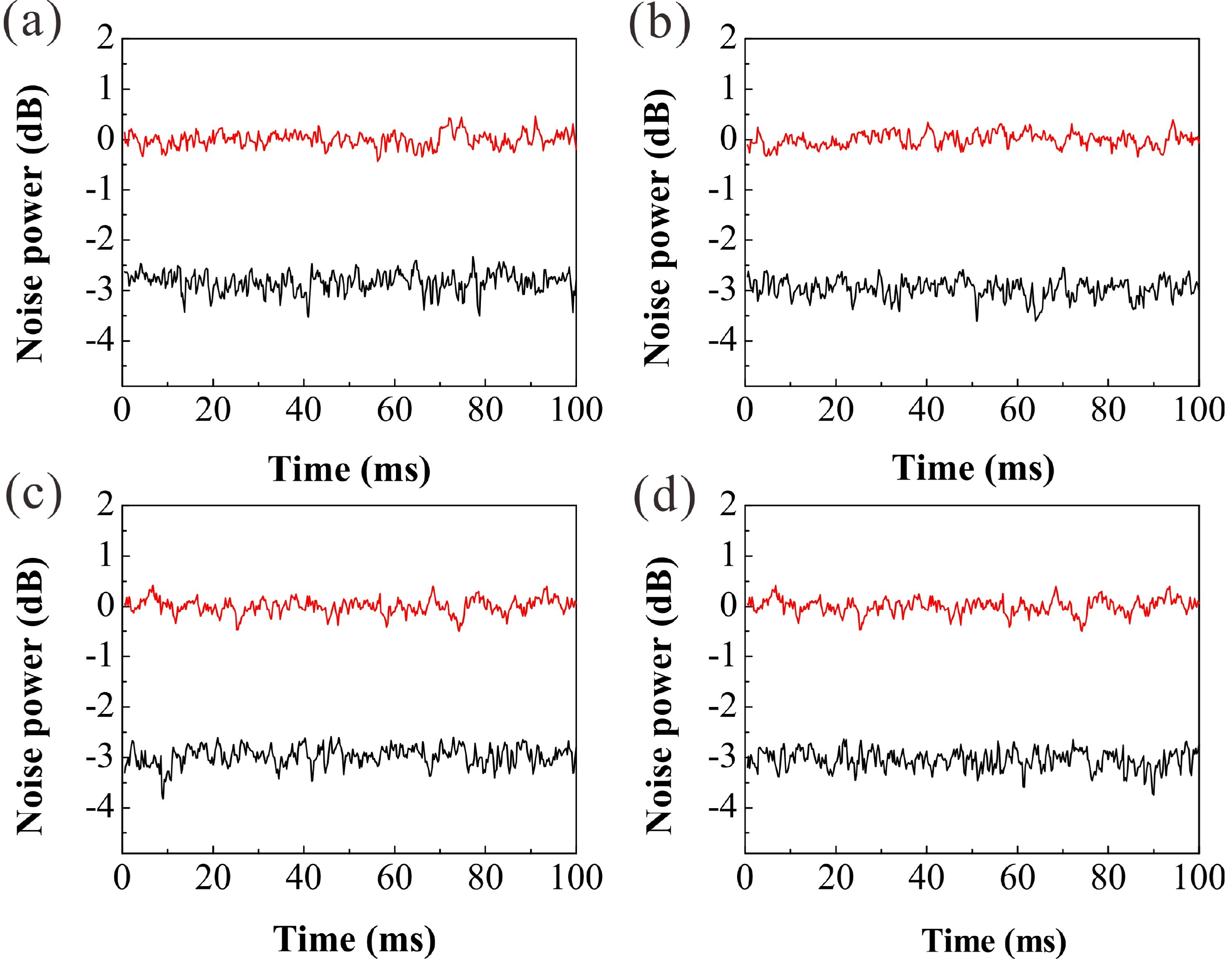}
\end{center}
\caption{The experimentally measured quantum correlation variances of the
original CV four-mode square Gaussian cluster state. Panels (a)--(d) show
the noise powers of $\Delta ^{2}\left( \hat{p}_{A}-\hat{x}_{C}-\hat{x}%
_{D}\right) $, $\Delta ^{2}\left( \hat{p}_{B}-\hat{x}_{C}-\hat{x}_{D}\right) 
$, $\Delta ^{2}\left( \hat{p}_{C}-\hat{x}_{A}-\hat{x}_{B}\right) $ and $%
\Delta ^{2}\left( \hat{p}_{D}-\hat{x}_{A}-\hat{x}_{B}\right) $,
respectively. The red and black lines are the normalized shot-noise-level
and correlated noise, respectively. The measurement frequency is 3 MHz, the
resolution bandwidth of the spectrum analyser is 30 KHz, and the video
bandwidth of the spectrum analyser is 300 Hz.}
\label{fig:S1}
\end{figure}

The four-mode entangled state used in the experiment is a continuous
variable (CV) square Gaussian cluster state of optical field at the sideband
frequency of 3 MHz and is prepared by coupling two phase-squeezed and two
amplitude-squeezed states of light on an optical beam-splitter network,\
which consists of three optical beam-splitters with transmittance of $%
T_{1}=1/5$ and $T_{2}=T_{3}=1/2$, respectively, as shown in Fig.~\ref%
{fig:scheme}(b) in the main text. Four input squeezed states are expressed
by 
\end{subequations}
\begin{align}  \label{eq:S1}
\hat{a}_{1}& =\frac{1}{2}[e^{r_{1}}\hat{x}_{1}^{(0)}+ie^{-r_{1}}\hat{p}%
_{1}^{(0)}],  \notag \\
\hat{a}_{2}& =\frac{1}{2}[e^{-r_{2}}\hat{x}_{2}^{(0)}+ie^{r_{2}}\hat{p}%
_{2}^{(0)}],  \notag \\
\hat{a}_{3}& =\frac{1}{2}[e^{-r_{3}}\hat{x}_{3}^{(0)}+ie^{r_{3}}\hat{p}%
_{3}^{\left( 0\right) }],  \notag \\
\hat{a}_{4}& =\frac{1}{2}[e^{r_{4}}\hat{x}_{4}^{(0)}+ie^{-r_{4}}\hat{p}%
_{4}^{\left( 0\right) }],
\end{align}%
where $r_{i}$ ($i=1,2,3,4$) is the squeezing parameter, $\hat{x}=\hat{a}+%
\hat{a}^{\dag }$ and $\hat{p}=(\hat{a}-\hat{a}^{\dag })/i$ are the amplitude
and phase quadratures of an optical field $\hat{a}$, respectively, and the
superscript of the amplitude and phase quadratures represent the vacuum
state. The transformation matrix of the beam-splitter network is given by 
\begin{equation}  \label{eq:S2}
U=\left[ 
\begin{array}{cccc}
-\sqrt{\frac{1}{2}} & -\sqrt{\frac{2}{5}} & -\frac{i}{\sqrt{10}} & 0 \\ 
\sqrt{\frac{1}{2}} & -\sqrt{\frac{2}{5}} & -\frac{i}{\sqrt{10}} & 0 \\ 
0 & \frac{i}{\sqrt{10}} & \sqrt{\frac{2}{5}} & -\sqrt{\frac{1}{2}} \\ 
0 & \frac{i}{\sqrt{10}} & \sqrt{\frac{2}{5}} & \sqrt{\frac{1}{2}}%
\end{array}%
\right] ,
\end{equation}
the unitary matrix can be decomposed into a beam-splitter network $%
U=F_{4}F_{3}I_{1}(-1)B_{34}(T_{3})F_{4}B_{12}(T_{2})B_{23}(T_{1})F_{3},$
where $B_{kl}(T_{j})$ stands for the linearly optical transformation on $j$%
th beam-splitter with transmission of $T_{j}$ ($j=1,2,3$), where $\left(
B_{kl}\right) _{kk}=\sqrt{1-T},\left( B_{kl}\right) _{kl}=\left(
B_{kl}\right) _{lk}=\sqrt{T},\left( B_{kl}\right) _{ll}=-\sqrt{1-T}%
~(k,l=1,2,3,4)$, are matrix elements of the beam-splitter. $F_{k}$ [$%
I_{k}(-1)$] denotes the $90%
{{}^\circ}%
$ ($180%
{{}^\circ}%
$) rotation in phase space of mode $k$, $\hat{a}_{k}\rightarrow i\hat{a}_{k}$
($\hat{a}_{k}\rightarrow -\hat{a}_{k}$). The output modes from the optical
beam-splitter network are expressed by 
\begin{align}  \label{eq:S3}
\hat{A}& =-\sqrt{\frac{1}{2}}\hat{a}_{1}-\sqrt{\frac{2}{5}}\hat{a}_{2}-i%
\sqrt{\frac{1}{10}}\hat{a}_{3},  \notag \\
\hat{B}& =\sqrt{\frac{1}{2}}\hat{a}_{1}-\sqrt{\frac{2}{5}}\hat{a}_{2}-i\sqrt{%
\frac{1}{10}}\hat{a}_{3},  \notag \\
\hat{C}& =i\sqrt{\frac{1}{10}}\hat{a}_{2}+\sqrt{\frac{2}{5}}\hat{a}_{3}-%
\sqrt{\frac{1}{2}}\hat{a}_{4},  \notag \\
\hat{D}& =i\sqrt{\frac{1}{10}}\hat{a}_{2}+\sqrt{\frac{2}{5}}\hat{a}_{3}+%
\sqrt{\frac{1}{2}}\hat{a}_{4},
\end{align}%
respectively. Here, we have assumed that four squeezed states have the
identical squeezing parameter ($r_{1}=r_{2}=r_{3}=r_{4}$). In experiments,
the requirement is easily achieved by adjusting the two NOPAs to operate
precisely at the same conditions. For our experimental system, we have
measured $r=0.345$. The quantum correlations between the amplitude and phase
quadratures are expressed by $\Delta ^{2}\left( \hat{p}_{A}-\hat{x}_{C}-\hat{%
x}_{D}\right) =\Delta ^{2}\left( \hat{p}_{B}-\hat{x}_{C}-\hat{x}_{D}\right)
=\Delta ^{2}\left( \hat{p}_{C}-\hat{x}_{A}-\hat{x}_{B}\right) =\Delta
^{2}\left( \hat{p}_{D}-\hat{x}_{A}-\hat{x}_{B}\right) =3e^{-2r}$, where the
subscripts correspond to different optical modes. Obviously, in the ideal
case with infinite squeezing ($r\rightarrow \infty $), these noise variances
will vanish and the better the squeezing, the smaller the noise terms.

According to the criteria for CV multipartite entanglement proposed by van
Loock and Furusawa \cite{Loock}, we deduce the inseparability conditions for
the CV four-mode square cluster state, which are 
\begin{eqnarray}  \label{eq:furusawa}
\Delta ^{2}\left( \hat{p}_{A}-\hat{x}_{C}-\hat{x}_{D}\right) +\Delta
^{2}\left( \hat{p}_{C}-\hat{x}_{A}-\hat{x}_{B}\right) &<&4,  \notag \\
\Delta ^{2}\left( \hat{p}_{A}-\hat{x}_{C}-\hat{x}_{D}\right) +\Delta
^{2}\left( \hat{p}_{D}-\hat{x}_{A}-\hat{x}_{B}\right) &<&4,  \notag \\
\Delta ^{2}\left( \hat{p}_{B}-\hat{x}_{C}-\hat{x}_{D}\right) +\Delta
^{2}\left( \hat{p}_{C}-\hat{x}_{A}-\hat{x}_{B}\right) &<&4,  \notag \\
\Delta ^{2}\left( \hat{p}_{B}-\hat{x}_{C}-\hat{x}_{D}\right) +\Delta
^{2}\left( \hat{p}_{D}-\hat{x}_{A}-\hat{x}_{B}\right) &<&4.
\end{eqnarray}%
When all the combinations of variances of nullifiers in the left-hand sides
of these inequalities are smaller than $4$ (which defines the normalized
boundary for inseparability, given a unit variance for each quadrature of
the vacuum state), then the four modes are in a fully inseparable CV square
cluster state.

The correlation variances measured experimentally are shown in Fig.~\ref%
{fig:S1}. They are $\Delta ^{2}\left( \hat{p}_{A}-\hat{x}_{C}-\hat{x}%
_{D}\right) =-2.84\pm 0.20$ dB, $\Delta ^{2}\left( \hat{p}_{B}-\hat{x}_{C}-%
\hat{x}_{D}\right) =-2.97\pm 0.19$ dB, $\Delta ^{2}\left( \hat{p}_{C}-\hat{x}%
_{A}-\hat{x}_{B}\right) =-2.97\pm 0.19$ dB and $\Delta ^{2}\left( \hat{p}%
_{D}-\hat{x}_{A}-\hat{x}_{B}\right) =-3.05\pm 0.19$ dB, respectively. From
these measured results we can calculate the combinations of the correlation
variances in the left-hand sides of the inequalities (\ref{eq:furusawa}),%
\footnote{%
Note that the experimental variances are measured in dB. To insert the
values into the inequalities~(\ref{eq:furusawa}), we need to convert them
back into dimensionless units, via the formula: $\Delta^2(\hat p_i - \hat
x_j - \hat x_k) = 3 \times 10^{\text{(variance in dB)}/10}$.} which are $%
3.07\pm 0.02$, $3.05\pm 0.02$, $3.03\pm 0.02$ and $3.01\pm 0.02$,
respectively. Thus all inequalities (\ref{eq:furusawa}) are simultaneously
satisfied, which confirms the prepared state is a fully inseparable CV
four-mode square cluster state.

\section{Measurement of the covariance matrix}

A Gaussian state is a state with Gaussian characteristic functions and
quasi-probability distributions on the multi-mode quantum phase space, which
can be completely characterized by its covariance matrix. The elements of
the covariance matrix are $\sigma _{ij}=Cov\left( \hat{\xi}_{i},\hat{\xi}%
_{j}\right) =\frac{1}{2}\left\langle \hat{\xi}_{i}\hat{\xi}_{j}+\hat{\xi}_{j}%
\hat{\xi}_{i}\right\rangle -\left\langle \hat{\xi}_{i}\right\rangle
\left\langle \hat{\xi}_{j}\right\rangle $, $i,j=1,2,\ldots ,8$, where $\hat{%
\xi}=(\hat{x}_{A},\hat{p}_{A},\hat{x}_{B},\hat{p}_{B},\hat{x}_{C},\hat{p}%
_{C},\hat{x}_{D},\hat{p}_{D})^{T}$ is a vector composed by the amplitude and
phase quadratures of four-mode states~\cite{Adesso2}. For convenience, the
covariance matrix of the original four-mode Gaussian state is written in
terms of two-by-two submatrices as 
\begin{equation}  \label{eq:S5}
\sigma =\left[ 
\begin{array}{cccc}
\sigma _{A\text{ }} & \sigma _{AB} & \sigma _{AC} & \sigma _{AD} \\ 
\sigma _{AB}^{T} & \sigma _{B\text{ }} & \sigma _{BC} & \sigma _{BD} \\ 
\sigma _{AC}^{T} & \sigma _{BC}^{T} & \sigma _{C\text{ }} & \sigma _{CD} \\ 
\sigma _{AD}^{T} & \sigma _{BD}^{T} & \sigma _{CD}^{T} & \sigma _{D}%
\end{array}%
\right] ,
\end{equation}

Thus the four-mode covariance matrix can be partially expressed as (the
cross correlations between different quadratures of one mode are taken as $0$%
) 
\begin{align}
\sigma _{A\text{ }}& =\left[ 
\begin{array}{cc}
\Delta ^{2}\hat{x}_{A} & 0 \\ 
0 & \Delta ^{2}\hat{p}_{A}%
\end{array}%
\right] ,  \notag \\
\sigma _{B\text{ }}& =\left[ 
\begin{array}{cc}
\Delta ^{2}\hat{x}_{B} & 0 \\ 
0 & \Delta ^{2}\hat{p}_{B}%
\end{array}%
\right] ,  \notag \\
\sigma _{C\text{ }}& =\left[ 
\begin{array}{cc}
\Delta ^{2}\hat{x}_{C} & 0 \\ 
0 & \Delta ^{2}\hat{p}_{C}%
\end{array}%
\right] ,  \notag \\
\sigma _{D\text{ }}& =\left[ 
\begin{array}{cc}
\Delta ^{2}\hat{x}_{D} & 0 \\ 
0 & \Delta ^{2}\hat{p}_{D}%
\end{array}%
\right] ,  \notag \\
\sigma _{AB\text{ }}& =\left[ 
\begin{array}{cc}
Cov\left( \hat{x}_{A},\hat{x}_{B}\right) & Cov\left( \hat{x}_{A},\hat{p}%
_{B}\right) \\ 
Cov\left( \hat{p}_{A},\hat{x}_{B}\right) & Cov\left( \hat{p}_{A},\hat{p}%
_{B}\right)%
\end{array}%
\right] ,  \notag \\
\sigma _{AC\text{ }}& =\left[ 
\begin{array}{cc}
Cov\left( \hat{x}_{A},\hat{x}_{C}\right) & Cov\left( \hat{x}_{A},\hat{p}%
_{C}\right) \\ 
Cov\left( \hat{p}_{A},\hat{x}_{C}\right) & Cov\left( \hat{p}_{A},\hat{p}%
_{C}\right)%
\end{array}%
\right] ,  \notag \\
\sigma _{AD\text{ }}& =\left[ 
\begin{array}{cc}
Cov\left( \hat{x}_{A},\hat{x}_{D}\right) & Cov\left( \hat{x}_{A},\hat{p}%
_{D}\right) \\ 
Cov\left( \hat{p}_{A},\hat{x}_{D}\right) & Cov\left( \hat{p}_{A},\hat{p}%
_{D}\right)%
\end{array}%
\right] ,  \notag \\
\sigma _{BC\text{ }}& =\left[ 
\begin{array}{cc}
Cov\left( \hat{x}_{B},\hat{x}_{C}\right) & Cov\left( \hat{x}_{B},\hat{p}%
_{C}\right) \\ 
Cov\left( \hat{p}_{B},\hat{x}_{C}\right) & Cov\left( \hat{p}_{B},\hat{p}%
_{C}\right)%
\end{array}%
\right] ,  \notag \\
\sigma _{BD\text{ }}& =\left[ 
\begin{array}{cc}
Cov\left( \hat{x}_{B},\hat{x}_{D}\right) & Cov\left( \hat{x}_{B},\hat{p}%
_{D}\right) \\ 
Cov\left( \hat{p}_{B},\hat{x}_{D}\right) & Cov\left( \hat{p}_{B},\hat{p}%
_{D}\right)%
\end{array}%
\right] ,  \notag \\
\sigma _{CD\text{ }}& =\left[ 
\begin{array}{cc}
Cov\left( \hat{x}_{C},\hat{x}_{D}\right) & Cov\left( \hat{x}_{C},\hat{p}%
_{D}\right) \\ 
Cov\left( \hat{p}_{C},\hat{x}_{D}\right) & Cov\left( \hat{p}_{C},\hat{p}%
_{D}\right)%
\end{array}%
\right] .
\end{align}

From the output modes given in Eq.~(\ref{eq:S3}) and the information of the
four input squeezed states given in Eq.~(\ref{eq:S1}), we can theoretically
obtain the amplitude and phase quadratures of the four-mode state and then
determine all the elements of the covariance matrix in Eq.~\ref{eq:S5}.
These are used for the theoretical predictions.

\begin{figure*}
\includegraphics[width=140mm]{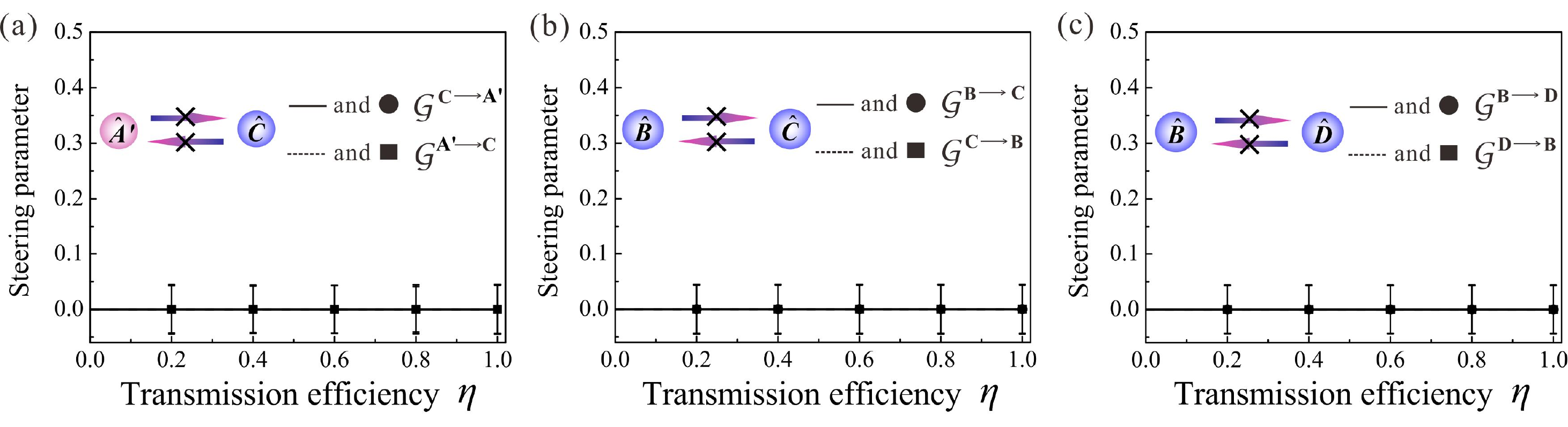}
\caption{Gaussian EPR steering between two modes of the cluster state,
supplementing Fig.~\protect\ref{fig:2modes} in the main text. (a)--(c)
Gaussian EPR steering between neighboring modes $\hat{A}^{\prime }$ and $%
\hat{C}$, $\hat{C}$ and $\hat{B}$, $\hat{B}$ and $\hat{D}$, respectively.
Clearly, no EPR steering is possible between these (1+1)-mode neighboring
modes in the CV four-mode square Gaussian cluster state under Gaussian
measurements. In all the panels, the quantities plotted are dimensionless.
The lines and curves represent theoretical predictions. The dots and squares
represent the experimental data measured at different transmission
efficiencies. Error bars represent $\pm $ one standard deviation and are
obtained based on the statistics of the measured noise variances.}
\label{fig:S2}
\end{figure*}
\medskip

\begin{figure*}
\includegraphics[width=170mm]{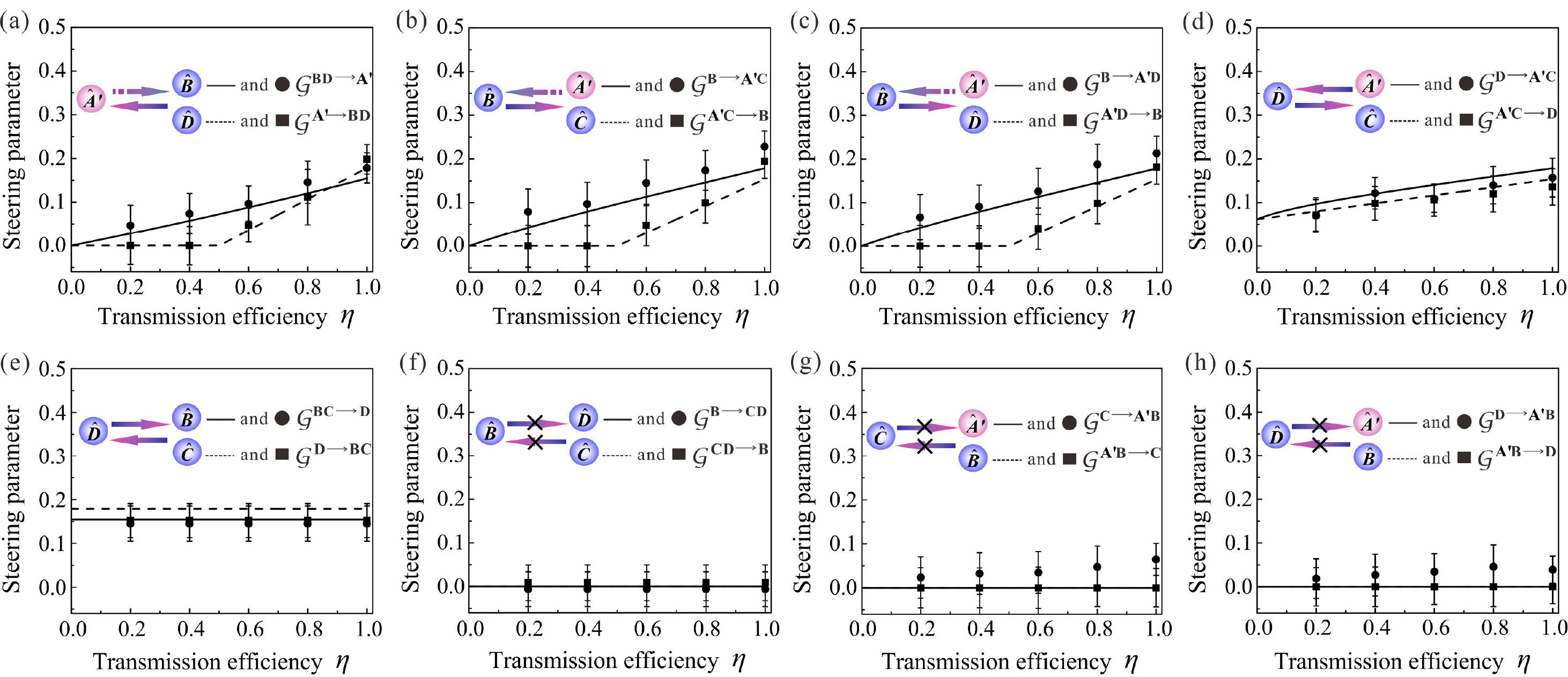}
\caption{Gaussian EPR steering between one and two modes of the cluster
state, supplementing Fig.~\protect\ref{fig:3modes} in the main text. (a)
One-way EPR steering between modes $\hat{A}^{\prime }$ and \{$\hat{B}$ and $%
\hat{C}$\} under Gaussian measurements. (b) One-way EPR steering between
modes $\hat{B}$ and \{$\hat{A}^{\prime }$ and $\hat{C}$\} under Gaussian
measurements. (c) One-way EPR steering between modes $\hat{B}$ and \{$\hat{A}%
^{\prime }$ and $\hat{D}$\} under Gaussian measurements. (d) $\hat{D}$ and \{%
$\hat{A}^{\prime }$ and $\hat{C}$\} can steer each other asymmetrically and
the Gaussian steerability grows with the increasing transmission efficiency.
(e) $\hat{D} $ and \{$\hat{B}$ and $\hat{C}$\} can steer each other
asymmetrically. (f) There is no EPR steering between $\hat{B}$ and \{$\hat{C}
$ and $\hat{D}$\} under Gaussian measurements. (g) There is no EPR steering
between $\hat{C}$ and \{$\hat{A}^{\prime }$ and $\hat{B}$\} under Gaussian
measurements. (h) There is no EPR steering between $\hat{D}$ and \{$\hat{A}%
^{\prime }$ and $\hat{B}$\} under Gaussian measurements. In all the panels,
the quantities plotted are dimensionless. The lines and curves represent
theoretical predictions. The dots and squares represent the experimental
data measured at different transmission efficiencies. Error bars represent $%
\pm $ one standard deviation and are obtained based on the statistics of the
measured noise variances.}
\label{fig:S3}
\end{figure*}

In the experiment, to partially reconstruct all relevant entries of the
associated covariance matrix of the state, we perform 32 different
measurements on the output optical modes. These measurements include the
amplitude and phase quadratures of the output optical modes, and the cross
correlations $\Delta ^{2}\left( \hat{x}_{A}-\hat{x}_{B}\right) $, $\Delta
^{2}\left( \hat{x}_{A}-\hat{x}_{C}\right) $, $\Delta ^{2}\left( \hat{x}_{A}-%
\hat{x}_{D}\right) $, $\Delta ^{2}\left( \hat{x}_{B}-\hat{x}_{C}\right) $, $%
\Delta ^{2}\left( \hat{x}_{B}-\hat{x}_{D}\right) $, $\Delta ^{2}\left( \hat{x%
}_{C}-\hat{x}_{D}\right) $, $\Delta ^{2}\left( \hat{p}_{A}-\hat{p}%
_{B}\right) $, $\Delta ^{2}\left( \hat{p}_{A}-\hat{p}_{C}\right) $, $\Delta
^{2}\left( \hat{p}_{A}-\hat{p}_{D}\right) $, $\Delta ^{2}\left( \hat{p}_{B}-%
\hat{p}_{C}\right) $, $\Delta ^{2}\left( \hat{p}_{B}-\hat{p}_{D}\right) $, $%
\Delta ^{2}\left( \hat{p}_{C}-\hat{p}_{D}\right) $, $\Delta ^{2}\left( \hat{x%
}_{A}+\hat{p}_{B}\right) $, $\Delta ^{2}\left( \hat{x}_{A}+\hat{p}%
_{C}\right) $, $\Delta ^{2}\left( \hat{x}_{A}+\hat{p}_{D}\right) $, $\Delta
^{2}\left( \hat{x}_{B}+\hat{p}_{C}\right) $, $\Delta ^{2}\left( \hat{x}_{B}+%
\hat{p}_{D}\right) $, $\Delta ^{2}\left( \hat{x}_{C}+\hat{p}_{D}\right) $, $%
\Delta ^{2}\left( \hat{p}_{A}+\hat{x}_{B}\right) $, $\Delta ^{2}\left( \hat{p%
}_{A}+\hat{x}_{C}\right) $, $\Delta ^{2}\left( \hat{p}_{A}+\hat{x}%
_{D}\right) $, $\Delta ^{2}\left( \hat{p}_{B}+\hat{x}_{C}\right) $, $\Delta
^{2}\left( \hat{p}_{B}+\hat{x}_{D}\right) $ and $\Delta ^{2}\left( \hat{p}%
_{C}+\hat{x}_{D}\right) $. The covariance elements are calculated via the
identities \cite{Steinlechner}%
\begin{align}
Cov\left( \hat{\xi}_{i},\hat{\xi}_{j}\right) & =\frac{1}{2}\left[ \Delta
^{2}\left( \hat{\xi}_{i}+\hat{\xi}_{j}\right) -\Delta ^{2}\hat{\xi}%
_{i}-\Delta ^{2}\hat{\xi}_{j}\right] ,  \notag \\
Cov\left( \hat{\xi}_{i},\hat{\xi}_{j}\right) & =-\frac{1}{2}\left[ \Delta
^{2}\left( \hat{\xi}_{i}-\hat{\xi}_{j}\right) -\Delta ^{2}\hat{\xi}%
_{i}-\Delta ^{2}\hat{\xi}_{j}\right] .
\end{align}

The steerability of Bob by Alice ($A \rightarrow B$) for a ($n_A+n_B$)-mode
Gaussian state under Gaussian measurements can be quantified by Eq.~(\ref%
{eqn:parameter}) in the main text, based on the symplectic eigenvalues
derived from the Schur complement of $A$ in the covariance matrix. In Figs.~%
\ref{fig:2modes}--\ref{fig:4modes} of the main text and Figs.~\ref{fig:S2}--%
\ref{fig:S3}, the lines and curves represent theoretical predictions based
on the theoretically calculated covariance matrix, while the dots and
squares report the measured steerability as evaluated from the
experimentally reconstructed covariance matrix.

\section{supplementary figures}

In this section, we provide additional figures that supplement the main
text. In particular, the additional experimental results of EPR steering
between neighboring modes $\hat{A}^{\prime }$ and $\hat{C}$, $\hat{B}$ and $%
\hat{C}$, $\hat{B}$ and $\hat{D}$ under Gaussian measurements are shown in
Fig.~\ref{fig:S2}, which supplements Fig.~\ref{fig:2modes} in the main text.
These figures support the result that no steering exist between neighboring
modes in the four-mode square Gaussian cluster entangled state under
Gaussian measurements.

We also provide the additional experimental results of EPR steering between
one and two modes [(1+2)-mode and (2+1)-mode partitions] of the CV four-mode
square cluster state under Gaussian measurements. The results of $\hat{A}%
^{\prime }$ and \{$\hat{B}$ and $\hat{C}$\}, $\hat{B}$ and \{$\hat{A}%
^{\prime }$ and $\hat{C}$\}, $\hat{B}$ and \{$\hat{A}^{\prime }$ and $\hat{D}
$\}, $\hat{D}$ and \{$\hat{A}^{\prime }$ and $\hat{C}$\}, $\hat{D}$ and \{$%
\hat{B}$ and $\hat{C}$\}, $\hat{B}$ and \{$\hat{C}$ and $\hat{D}$\}, $\hat{C}
$ and \{$\hat{A}^{\prime }$ and $\hat{B}$\}, $\hat{D}$ and \{$\hat{A}%
^{\prime }$ and $\hat{B}$\} are shown in Fig.~\ref{fig:S3}, which
supplements Fig.~\ref{fig:3modes} in the main text. All the results provide
complete support to our analysis and conclusions as discussed in the main
text.

\end{document}